\documentclass[
superscriptaddress,
nofootinbib,
amsmath,amssymb,
longbibliography,
twocolumn 
]{revtex4-2}

\usepackage{wrapfig}
\usepackage{graphicx}
\usepackage{lineno}
\usepackage[utf8]{inputenc}
\usepackage{rsfso}
\usepackage{amsmath,color}
\usepackage{graphicx}
\usepackage{dcolumn}
\usepackage{bm}
\usepackage{amsfonts}
\usepackage{amsthm}
\usepackage{breqn}
\usepackage{xcolor}
\newcommand{\ML}[1]{\textcolor{black}{#1}}
\newcommand{\MLL}[1]{\textcolor{black}{#1}}
\newcommand{\MLrev}[1]{\textcolor{black}{#1}}
\newcommand{\MLreV}[1]{\textcolor{black}{#1}}
\newcommand{\MLreVV}[1]{\textcolor{black}{#1}}
\newcommand{\MLeditor}[1]{\textcolor{black}{#1}}
\newcommand{\MLeditoR}[1]{\textcolor{black}{#1}}

\usepackage{hyperref}
\begin{document}
\title{Symmetries at the Anderson transition \\ of correlated two-dimensional electrons}

\title{Anderson transition symmetries at the band-edge \\ of a correlated Sn/Si monolayer}

\author{Mathieu Lizée}
\email{lizee@fhi-berlin.mpg.de}
\thanks{Current address: Fritz Haber Institute of the Max Planck Society, Berlin, Germany}
\affiliation{Laboratoire de Physique de l'Ecole Normale Supérieure - ENS Université PSL - Paris France}
\affiliation{Sorbonne Université - CNRS - Institut des NanoSciences de Paris - UMR7588 - F75252 Paris - France}

\author{Mohammadmehdi Torkzadeh}
\author{François Debontridder}
\author{Marie Hervé}
\author{Christophe Brun}
\affiliation{Sorbonne Université - CNRS - Institut des NanoSciences de Paris - UMR7588 - F75252 Paris - France}

\author{Igor Burmistrov}
\affiliation{Landau Institute for Theoretical Physics - Chernogolovka - Russia}

\author{Tristan Cren}
\email{tristan.cren@sorbonne-universite.fr}
\affiliation{Sorbonne Université - CNRS - Institut des NanoSciences de Paris - UMR7588 - F75252 Paris - France}


\begin{abstract}

	\MLeditor{Anderson localization is predicted to enhance the critical temperature of disordered superconductors. Despite a huge body of theoretical work based on non-linear sigma models, experiments are lacking to understand correlated electrons in disordered potentials. In this study, we investigate a tin monolayer on silicon, a material known for its likely antiferromagnetic Mott-correlated groundstate. We analyze the statistical properties of tunneling conductance maps of increasingly localized states as we approach the edge of the valence band. Using multifractal analysis, we show that the system follows an exact symmetry relation based on the algebraic structure of nonlinear sigma-models (NL$\sigma$Ms). We anticipate that this symmetry may be broken in specific -- e.g. chiral -- electronic phases. Finally, we point out that multifractal analysis can equally be applied to universal conductance fluctuations in magneto-transport experiments, thus providing a powerful tool to probe the underlying symmetries of disordered electronic phases.}

\end{abstract}

\maketitle



\section*{Introduction}

The combined role of disorder and electronic interactions is crucial in the context of low-dimension superconductivity where the multifractality of wave functions close to the localization transition is believed to enhance the critical temperature $T_{\rm c}$ \cite{sacepe2008disorder,feigel2010fractal,burmistrov2012enhancement}. Strikingly at odds with the usual quantum criticality picture, recent kinetic inductance measurements in indium-oxide films hint at a first-order transition from a superconductor to a Bose glass \cite{charpentier2025first}. In normal metals on the other hand, the interplay of Coulomb repulsion with disorder yields the Mott-Anderson transition \cite{belitz1994anderson,knyazev2008metal}: in 2D systems, the Coulomb repulsion is predicted to hinder localization, rendering metallic an otherwise localized phase \cite{burmistrov2013multifractality}. The possibility of non-trivial topology in a strongly disordered band insulator -- the so-called topological Anderson insulator -- remains an open question \cite{groth2009theory}.

\MLrev{Our understanding of such interacting mesoscopic conductors strongly relies on non-linear $\sigma$-models (NL$\sigma$Ms), a class of field-theories introduced in the context of weak localization \cite{efetov1980interaction} and later generalized to interacting electronic systems \cite{finkelstein1983frequency}. NL$\sigma$M describe the low energy excitations of disordered metals in terms of diffusons and cooperons but can be renormalized to very high disorder, at the Anderson transition. Although their domain of validity is not known, they are widely used to describe metal-insulator transitions and have even predicted multifractally-enhanced superconductivity in the strong interaction -- strong disorder limit \cite{feigel2010fractal,burmistrov2012enhancement,andriyakhina2022multifractally}. In this work, we take advantage of the so-called Weyl-group symmetry, deeply rooted in the mathematical structure of NL$\sigma$Ms, to test their validity across the effective metal-insulator transition at the band-edge of a correlated 2d system \cite{mirlin1994distribution,mirlin2006exact,gruzberg2011symmetries,gruzberg2013classification}}

In recent years, tunneling local density of states (LDOS) maps have revealed critical scalings and fractal-like structures in the quantum Hall effect regime \cite{hashimoto2008quantum}, at the surface of a \MLrev{dilute magnetic semiconductor} \cite{richardella2010visualizing} and close to the band edges of monolayer MoS$_2$ \cite{jack2021visualizing,shin2023characterizing,shin2023structural}. For superconductors, a granularity of the gap width order parameter was observed in relatively strongly disordered thin films  \cite{ghosal2001inhomogeneous,sacepe2008disorder,carbillet2016confinement,ghosal2001inhomogeneous,carbillet2020spectroscopic} close to the regime where NL$\sigma$Ms predict $T_{\rm c}$-enhancement by Anderson localization \cite{feigel2010fractal,burmistrov2012enhancement,andriyakhina2022multifractally}. At lower disorder, enhanced fluctuations of LDOS close to coherence peaks were observed \cite{brun2014remarkable,rubio2020visualization,stosiek2021multifractal,lizee2023local} and rationalized as a real-space analog of universal conductance fluctuations (UCF) \cite{lizee2023local}. \ML{These high spectral resolution tunneling maps provide unmatched insights on the role of disorder in low-dimensional electronic systems.}


\begin{figure*}[!ht]
	\centering
	\includegraphics[width=\linewidth]{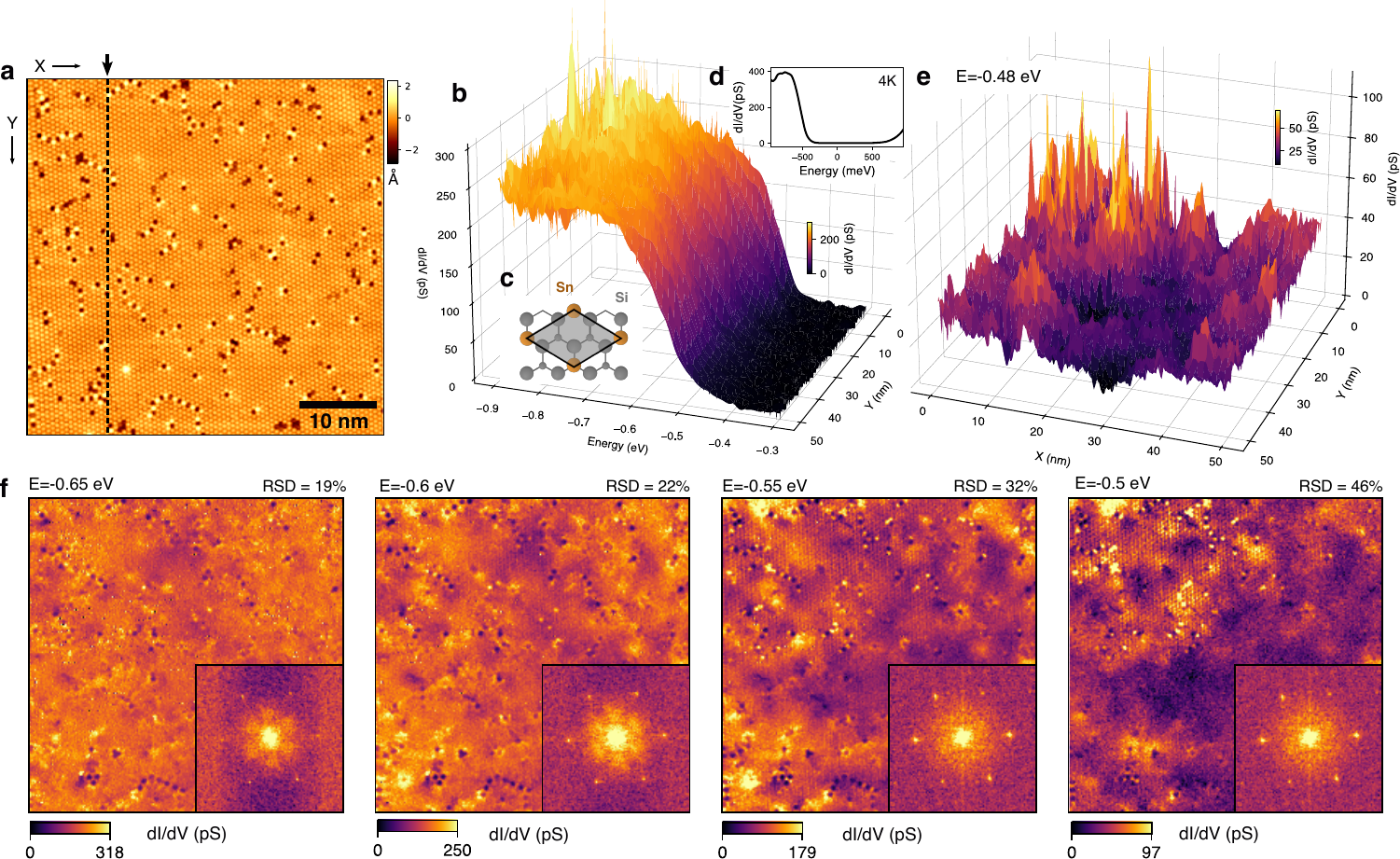} 
	\caption{\MLeditor{\textbf{Local-density of state maps through the valence-band edge of the Sn/Si monolayer}}\label{Figure1_3d_maps} \textbf{a} Topographic map of the tin monolayer deposited on silicon. \MLrev{For both topography and spectroscopic data, the bias voltage is -1V and the current setpoint 100 pA. Except is specified otherwise, the temperature is fixed at 300 mK throughout the paper.} \textbf{b} dI/dV spectra taken at the lower band-edge of the tin monolayer along a vertical line (cf. the black arrow indicates the X position on \textbf{a}). The color bar codes for the dI/dV values. \textbf{c} Crystalline structure of the tin $\sqrt{3}\times \sqrt{3}$ tin reconstruction on a Si(111) surface. 
	\textbf{d} dI/dV spectroscopy at 4K reveals a large gap of roughly 0.6 eV. \textbf{e} dI/dV map at -0.48 eV, very close to the band-edge. \MLeditor{The colour bar codes for the differential conductance at a given energy and position in pS.} \textbf{f} dI/dV maps at fixed energy at E = $\{$ -0.65, -0.6, -0.55, -0.5 $\}$ eV along with their Fourier transform \MLeditor{(quasi-particle interference patterns). The Relative Standard Deviation (RSD) of the dI/dV map is given on each panel and the colour bar codes for the differential conductance dI/dV.}}
\end{figure*}

Due to its 2D nature and to the closeness in energy of exchange interaction and on-site Coulomb repulsion, the 1/3 monolayer $\sqrt{3}\times\sqrt{3}$ reconstruction of tin on silicon hosts very rich electronic phases. At 77K, a transfer of spectral weight from the Fermi level to two Hubbard bands, as well as the appearance of a quasiparticle peak \MLrev{upon hole-doping}, shows its Mott insulating behavior \cite{ming2017realization,li2013magnetic,modesti2007insulating}. Because of its triangular lattice, it could realize the spin 1/2 triangular antiferromagnetic Heisenberg model which hosts quantum spin liquids and chiral spin states. A possible chiral superconductivity with a first hint of edge channels was recently observed in the p-doped monolayer \cite{ming2023evidence}.

In this study, we investigate the $\sqrt{3}\times\sqrt{3}$ reconstruction of tin on n-doped silicon with high-resolution Scanning Tunneling Spectroscopy (STS). We find the mobility edge using spatial correlation functions and measure energy-dependent multifractal scalings across the metal-insulator transition. Then, we use this model system to test NL$\sigma$Ms symmetry \cite{mirlin1994distribution,mirlin2006exact,gruzberg2011symmetries} in the high interaction-high disorder regime where their breakdown may be expected. 

Interestingly, beyond spatial LDOS maps, our approach could be applied to magneto-transport experiments \MLrev{since UCF \cite{gruzberg2013classification,lerner1988distribution}} also show multifractal statistics. \MLreV{Multifractal scalings of magneto-conductance were reported for high-mobility graphene transistors as functions of Fermi level and temperature \cite{amin2018exotic}. We anticipate from our results and the extensive body of theoretical literature on scaling-symmetry correspondence \cite{evers2008anderson} that multifractal spectra of UCF could become a key tool in the exploration of quantum phases, far beyond the study of Anderson localization.} 

\section*{Results}
\MLeditor{
We prepared the $\sqrt{3}\times\sqrt{3}$ reconstruction of tin on silicon by evaporating 1/3 monolayer of tin on a n-doped Si (111) crystal with a resistivity in the m$\Omega$.cm range at room temperature, following a well established procedure \cite{modesti2007insulating,ming2017realization, li2013magnetic} (see Methods).} We obtained an almost pure $\sqrt{3}\times\sqrt{3}$ reconstruction of tin on silicon (111) with domains larger than 100 nm and roughly 3 $\%$ of Si-substitutional and Sn-vacancy defects (see Figure \ref{Figure1_3d_maps}\textbf{a}). We also report small domains of the denser $2 \sqrt{3} \times 2 \sqrt{3}$ phase, in agreement with previous studies \cite{yi2018atomic}  -- see Supplementary Methods and Supplementary Figure 1. \MLrev{This system has been extensively studied theoretically, with strong efforts to account for Coulomb repulsion and exchange interactions \cite{profeta2007triangular,hansmann2013long,schuwalow2010realistic,li2013magnetic,lee2014antiferromagnetic}. Although the nature of its ground-state remains debated, a consensus is reached on its 2D nature close to the Fermi level, with bulk silicon bands being at least 250 meV away from the surface band at any point in the Brillouin zone. Note that the first silicon layers nevertheless have an important contribution in surface states \cite{schuwalow2010realistic,lee2014antiferromagnetic}.}

Now, focusing on the ground state, we measure tunneling spectroscopic maps at 300 mK across the edge of the \MLrev{valence} band ([-1,-0.3] V), see \ML{the} line cut in Figure \ref{Figure1_3d_maps}\textbf{b}. The tunneling spectrum at 4 K (Figure \ref{Figure1_3d_maps}\textbf{d}) reveals a clear insulating behavior with a large gap of about 0.6 eV.  Spin-resolved ARPES measurements \cite{jager2018alpha} and our hybrid-functionals DFT calculations (not yet published) show that this low temperature phase \MLrev{-- which replaces the Mott insulator reported by Ming et al.\cite{ming2017realization} below roughly 20 K --} probably corresponds to a row-wise antiferromagnetic ground state and is due to a subtle interplay of strong on-site Coulomb repulsion and \MLrev{exchange interactions mediated by the first silicon layers}. 

\begin{figure*}[!ht]
	\centering
	\includegraphics[width=\linewidth]{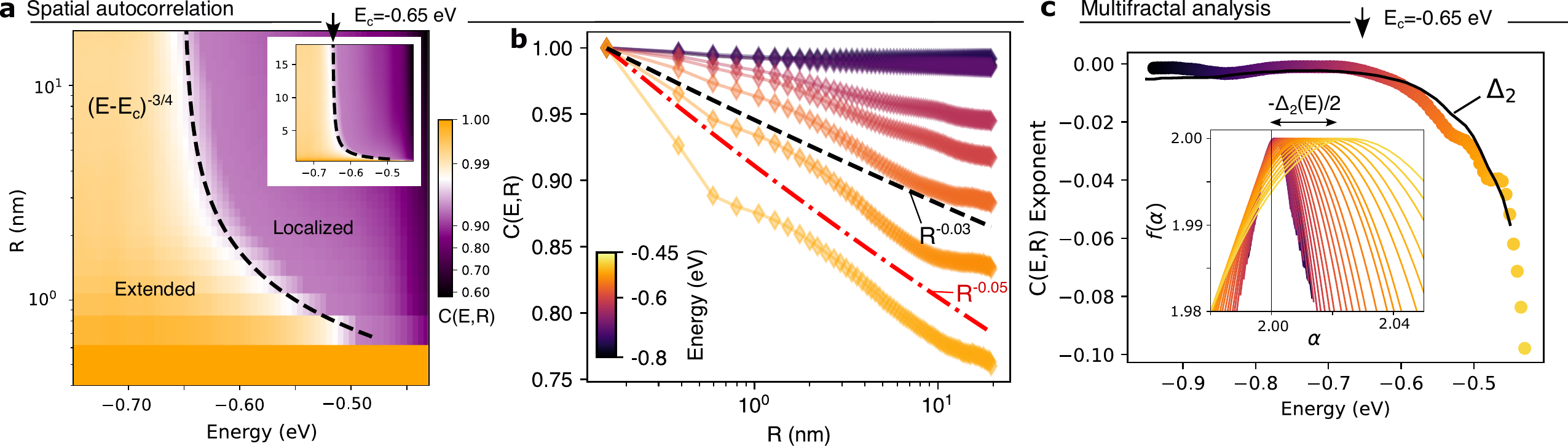} 
	\caption{\label{Correlation_heat_maps}\MLeditor{\textbf{Local density of states spatial correlation functions}} \textbf{a} Iso-energy radial correlation functions \MLrev{C(E,R)} of dI/dV maps at fixed energy. The x-axis denotes the energy of the maps and the y-axis the log of distance. The inset represents the same data but with a linear R axis. The black dashed line is a power law $\vert E-E_{\rm c}\vert^{-\nu_{\rm app}}$ with $E_{\rm c}=-0.65$ eV and $\nu_{\rm app} = 0.75$. \textbf{b} \ML{Radial correlations as function of energy, with E ranging from -0.8 eV to -0.45 eV.}  The black and red lines follow power laws of exponents -0.03 and -0.05. \MLeditor{The colour bar accounts for the energy of the LDOS map.} \textbf{c} \MLrev{Power law exponent of $C(E,R)$ fits as a function of energy plotted along with the fractal scaling exponent $\Delta_2$ measured by multifractal analysis.} The inset shows multifractal spectra taken at all energies \MLrev{(zooming on the region where $f(\alpha) \sim 2$)}. The color codes for energy as on the main plot. This panel shows that the exponent of correlation functions fitted from panel \textbf{b} perfectly correspond to fractal exponent $\Delta_2$, as predicted by theory \cite{burmistrov2013multifractality}.}
\end{figure*}

Deep in the \MLrev{valence} band (Figure \ref{Figure1_3d_maps} \textbf{e-f}), we observe \MLrev{extended} states, hereafter called metallic, with relatively low dispersion around the mean value (relative standard deviation (RSD) of $19\%$ at $-0.65$ eV). Close to the gap-edge on the other hand, the density of states is extremely inhomogeneous and concentrates in small regions of the sample whose shape is uncorrelated with the disorder's distribution, a clear hallmark of Anderson localization (RSD of $46\%$ at $-0.5$ eV). The Fourier transform of the conductance maps of these states show neat Bragg peaks of the $\sqrt{3}\times\sqrt{3}$ reconstruction. At \MLrev{higher binding energies}, a well-developed quasiparticle interference (QPI) signal reflects the band structure of the material and shows the delocalized nature of these electronic states. By contrast, when approaching the band edge, the QPIs become fuzzy while states appear increasingly localized in the real space. A detailed analysis of these QPI patterns will be published elsewhere. \MLreVV{In this work, we focus on understanding the spatial and spectral structure of LDOS maps close to the band edge.}

\subsection*{Metal -- insulator transition}
Critical scalings of LDOS distributions close to metal-insulator transitions were previously reported at the surface of a \MLrev{diluted magnetic semiconductor} \cite{richardella2010visualizing} and at the band edge of a monolayer transition-metal dichalchogenides \cite{shin2023structural}. On the theoretical side, a power-law decay of LDOS correlations is expected close to the metal-insulator transition \cite{burmistrov2013multifractality}. Let us then plot the angle-averaged 2-point correlation function of $\eta=dI/dV$ on Figure \ref{Correlation_heat_maps} \textbf{a-b}: 
\begin{equation}\label{spatial_corr}
	C(E,R) = \frac{\langle \eta(E,r) \eta(E,R+r) \rangle_r}{\langle \eta(E) \rangle^2}
\end{equation}
where $\langle \rangle_r$ denotes spatial averaging. On the $C(E,R)$ map (see \textbf{a-b}), \MLreV{we find that states at energies $E<-0.65$ eV keep very high spatial correlations at long distance whereas those at energies $E>-0.65$ show a strong decay with R. In line with similar observations on MoS$_2$ \cite{shin2023structural}, we rationalize this extended-to-localized crossover as an effective metal-insulator transition with $E_{\rm c} = -0.65$ eV for the mobility edge.} On panel \textbf{a}, the black dashed line follows a power law $\vert E-E_{\rm c}\vert^{-\nu_{\rm app}}$ yielding an \MLrev{apparent} critical exponent $\nu_{\rm app} = 0.75 \pm 0.1$. \MLreV{Surprisingly, $\nu_{\rm app} = 0.75$ does not respect the Harris criterion for the correlation length exponent of phase transitions in disordered systems $\nu > 2/d=1$ \cite{harris1974effect}. Although this suggests that $\nu_{\rm app}$ may not correspond to a true critical exponent and despite the extended character of states for $E<E_{\rm c}$ remaining uncertain, we still use the concepts of 'extended states' and 'mobility edge' in the following, since they make perfect sense at the scale of our STM map.}

On panel \textbf{b}, \ML{we observe} that the decay of correlations with distance is \ML{compatible with a power-law close to the band-edge as pointed by the black and red lines that have respective exponents $-0.03$ and $-0.05$.} \MLrev{In a diffusive system \MLreVV{where the spatial extent of electronic interference is set by the Thouless length $L_{\rm Th}$, one expects spatial correlations to follow} $C(E,R) \sim (L_{\rm Th}/R)^{-\Delta_2}$ with the exponent $\Delta_2$ being the anomalous fractal exponent, defined \MLreVV{such that for the density of states $\rho$,} $\langle\rho^2\rangle_r/\langle\rho\rangle_r^2 \sim r^{-\Delta_2}$ \cite{burmistrov2013multifractality}}. \MLreVV{Note that throughout the paper, we write $\rho$ for the density of states and $\eta$ for the differential tunneling conductivity}. $\Delta_2$ can be obtained independently of correlation functions through the multifractal analysis of LDOS maps (see panel \textbf{c}), as was experimentally reported on a handful of systems \cite{richardella2010visualizing,zhao2019disorder,rubio2020visualization,jack2021visualizing,shin2023structural}. From the radial correlation functions plotted on Figure \ref{Correlation_heat_maps}\textbf{b}, we extract the power-law exponents of $C(E,R)$ and plot them against energy on panel \textbf{c}. Using multifractal analysis, we compute the spatial scaling of the LDOS's second moment $\rho^2$ and extract the multifractal exponent $\Delta_2$, which we also plot on Figure \ref{Correlation_heat_maps}\textbf{c}. Clearly, $\Delta_2$ matches the power-law exponent of $C(E,R)$ at all energies, as predicted theoretically \cite{burmistrov2013multifractality}.

\paragraph*{Multifractal analysis}
\MLrev{Multifractal analysis generalizes traditional fractal analysis to the scaling dimension of all q$^{\rm th}$-moments $\tilde{\rho}^q$ of the LDOS. This set of scaling dimension can be Legendre-transformed into the multifractal spectrum $f(\alpha)$, \MLreVV{showing a typical inverted parabolic shape. The $f(\alpha)$ spectrum can be thought of as a distribution function of regions with critical exponent $\alpha$, or more precisely as the fractal dimension of the set of regions with scaling dimension $\alpha$ \cite{chhabra1989direct,schreiber1991multifractal} (see \MLeditor{Supplementary Note 1}). As a consequence, the criticality spectrum of non-fractal states is typically an infinitely narrow peak (a Dirac distribution) centered at the set's dimension (here d=2). In contrast, the spectrum of monofractal states is also a Dirac distribution but centered at the set's fractal dimension (d<2). More generally, multifractal states show a range of scaling dimensions which is described by a broadened distribution $f(\alpha)$.} As they can be computed from NL$\sigma$Ms \cite{evers2008anderson}, tight-binding Anderson models \cite{schreiber1991multifractal,stosiek2021multifractal} and experimentally measured from STM maps \cite{richardella2010visualizing,rubio2020visualization,jack2021visualizing} but also \MLrev{magneto-}transport data \cite{amin2018exotic}, multifractal spectra have become a central tool in the study of mesoscopic conductors. In the inset of Figure \ref{Correlation_heat_maps}\textbf{c}, we plot the multifractal spectra $f(\alpha)$ for various energies. Note that spectra for increasingly localized states are broadened and shifted to higher $\alpha$ values, the shift's amplitude being also given by the 2$^{\rm nd}$ scaling exponent $\alpha_0 -2 = - \Delta_2/2$. Clearly, $\Delta_2$ strongly increases in magnitude across the metal-insulator transition, from roughly $-10^{-3}$ in the metallic regime up to $-10^{-1}$ in the localized phase, suggesting a Anderson-localization origin of the increased multifractality.} \MLrev{This increased multifractality at the band edge is in very good agreement with previous measurements on Pb-Bi surface alloy \cite{jack2021visualizing}} and MoS$_2$ \cite{shin2023structural}.

\subsection*{Energy-scaling of LDOS fluctuations}
Let us now probe LDOS fluctuations on the localized side of the transition, between the mobility edge $E_{c} = -0.65$ eV and the band-edge. We correct for tip-height fluctuations by normalizing the differential conductance at set-point energy $E_\Lambda$ (see \MLeditor{Supplementary Methods}) and plot the \MLrev{spatially-averaged conductance} $\eta_{\Lambda}(E)$ on Figure \ref{Figure_scalings} \textbf{a}. On panel \textbf{b}, we show the LDOS normalized variance
\begin{equation}\label{def_sigma}
	\sigma^2(E) = \frac{\langle \delta \eta_{\Lambda}^2(E) \rangle_r}{\langle \eta_{\Lambda}(E)\rangle_r^2}
\end{equation}
By fitting the vanishing of mean-density of states, we find the band-edge to be at $E_{\rm edge} = -0.38 \pm 0.01$ eV (\textit{cf.} \textbf{a}-inset). Between this threshold and $-0.45$ eV, we have a high-noise region where a substantial proportion of pixels shows negative tunneling conductance and which we discard in further statistical analysis. Clearly, whereas the density of states decreases in a roughly linear way, $\sigma^2$ rises very sharply on almost two decades (\textbf{b}).

\MLrev{To perform a scaling analysis, we} shift the origin of energies to the band-edge $E_{\rm edge} = -0.38$ eV and compute the normalized variance of dI/dV maps at energy $E$ as a function of their distance to the gap $\omega = \vert E - E_{\rm edge} \vert$. As shown on Figure \ref{Figure_scalings} \textbf{c}, we obtain a scaling law $\sigma_{\rm exp}^2(E) \sim \omega^{-1.7}$ suggesting a critical behavior close to the band edge. We reproduced this measurement at the exact same position with a 6.5 Tesla magnetic field and found that this scaling law to slightly depart from the zero-field case with $\omega^{-1.9}$ instead of $\omega^{-1.7}$. \MLeditor{In Supplementary Note 2, Supplementary Figure 4}, we show that LDOS maps at zero and 6.5 T are extremely similar, as we can expect for an antiferromagnetic Mott-exchange insulator with an on-site Coulomb energy of about 1 eV. Finally, we report a log-scaling relation between the mean and variance of LDOS maps:
$\sigma^2 = -0.22 \textrm{log} \langle \eta_{\Lambda} \rangle_r/320$ pS
holding on more than a decade (\textit{cf.} \textbf{d}). We are not aware of such a scaling being reported before and believe it is valuable observation on the role of density of states on Anderson transitions. \MLrev{Let us mention here that these scaling laws are empirical observations on roughly one decade and do not suffice to \MLreVV{fully} demonstrate a critical behavior.}

\paragraph*{Numerical exploration}
To rationalize these scalings, we used the prototypical numerical model of a disordered metal: the Anderson tight-binding model on a square lattice for varying disorder strength. We found the band-edge scalings in to be relatively good agreement with the experiments. Considering the possibility that the ground state might have antiferromagnetic order, we intentionally broke time-reversal symmetry using a random Peierls field in each lattice cell instead of an on-site disorder. In the framework of Wigner-Dyson classification \cite{dyson1962statistical}, we expect this model to be of unitary class in contrast with the orthogonal class of the Anderson model. Overall, we find a slightly better match for the unitary model than for Anderson model, but the moderate agreement with experiments is insufficient to unambiguously tip the balance towards one or the other symmetry class\MLeditor{(see Supplementary Note 3 and Supplementary Figure 5 for details)}.

\subsubsection*{Estimating $g(E)$}

As an alternative to tight-binding numerical models, we used analytical results from the non-linear $\sigma$ models to extract the energy-dependent effective disorder strength. In this framework, we expect $\sigma^2$ (and the scaling exponent $\Delta_2$) to scale like the inverse dimensionless conductance of the layer $g = G_{\Box} h/e^2$  \cite{burmistrov2014tunneling} -- for the same reasons that the weak localization correction to conductivity scales like $1/g$. To \MLreVV{test this scaling, we estimate the energy-dependent conductivity $g(E)$} across the metal-insulator transition. \MLreVV{To do so, we used the minimal model for a 2D metal of a diffusive parabolic band, with its edge at $E_{\rm edge} = -0.38$ eV.}

\paragraph*{Parabolic band Drude model}
In the diffusive regime, we write $g(E)=\hbar D(E) \rho(E)$ where $\rho$ is the density of states and the diffusion coefficient depends on the electronic velocity at energy E and the elastic scattering rate $D(E)=v(E)^2/\gamma_{\rm el}$. Note that $\gamma_{\rm el}$ is energy independent at first order because charge carrier velocity $\sim \hbar k/m$ compensates the scattering cross-section $\sim 1/k$. Inferring the elastic scattering rate from the density of defects in the topographic map, we find $\gamma_{\rm el} \sim 170$ THz. We estimate the energy dependent density of states from a parabolic approximation ($\rho_0 \sim 2m/2\pi\hbar^2$) and the experimental dI/dV curve (see \MLeditor{Supplementary Note 4 and Supplementary Figure 6} for details). \MLreVV{From this model, we obtain the energy-dependent dimensionless conductance $g(E)$ which we compare to LDOS variance $\sigma^2$ to test the theoretical prediction that $\sigma^2(E) \sim 1/4\pi g(E)$.}

\begin{figure*}[!ht]
	\centering
	\includegraphics[width=\linewidth]{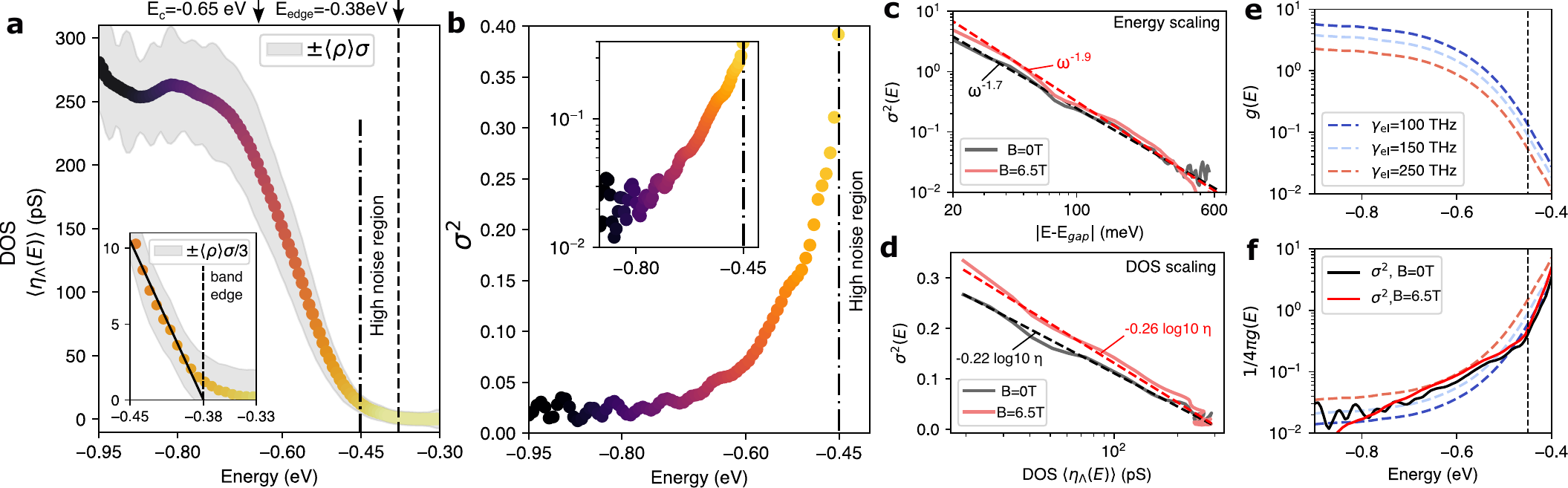}
	\caption{\label{Figure_scalings} \MLeditor{\textbf{Energy-dispersion of the local density of states variance}} \textbf{a} Mean normalized dI/dV spectrum $\langle\eta_{\Lambda}(E) \rangle_r$ as a function of energy. The grey area denotes the standard deviation on the map. For $E>-0.45$ eV, a large part of the grid's pixels have negative dI/dV, corresponding to a dominance of noise in this highly insulating region. We shall therefore focus on the $E<-0.45$ eV region. \textbf{b} Lin-lin and lin-log plots of \MLeditor{the normalized variance computed on an iso-energy LDOS map $\sigma^2 = \langle\delta \rho^2\rangle/\langle \rho \rangle^2$.} On \textbf{c}, we shift the origin of energies to the band edge and show that adding a magnetic field of 6.5 T slightly changes the scaling law. \textbf{d} Variance $\sigma^2(E)$ \textit{versus} mean $\langle \eta_{\Lambda}(E)\rangle_r$ of normalized tunneling conductance. \textbf{e} Dimensionless conductance $g(E)$ in the parabolic diffusive band model ($g(E)=\hbar \rho(E) D(E)$) as a function of energy, for different elastic scattering rates $\gamma_{\rm el}$. The Thouless energy is set at $E_{\rm Th} = 30 \mu$eV. \textbf{f} Comparison of the experimental $\sigma^2(E)$ with the weak-disorder model $1/4\pi g(E)$ for $\gamma_{\rm el} =\{100,150,250\}$ THz.}
\end{figure*}

\paragraph*{Comparison with $\sigma^2(E)$}
On Figure \ref{Figure_scalings} \textbf{e}, we plot the energy-dependent dimensionless conductance $g(E)$ for $\gamma_{\rm el} \in\{100, 150, 250\}$ THz, assuming the Thouless energy to be of the order of the thermal energy $E_{\rm Th} \sim k_BT =30$ $\mu$eV. On panel \textbf{f}, we plot $1/4\pi g(E)$ along with the experimental $\sigma^2$ curve. We find a very good agreement, suggesting that the elastic scattering rate is indeed very close to our estimate of 170 THz. With an electronic velocity of the order of 3$\cdot$10$^5$ m/s in the extended regime, this corresponds to a mean free path of about 2 nm, consistent with the defect density in the topographic map. From this experimental agreement, we can be relatively confident in the energy-dependent effective conductivity of panel \textbf{e} : between -0.9 and -0.45 eV, $g(E)$ drops by almost 2 orders of magnitude between roughly 5 and 0.1. \MLreVV{Let us reflect here on the fact that our analysis allows inferring $g(E)$, an energy-dependent dynamic quantity involving scattering rates, from equilibrium interference patterns.}



\subsection*{Non-linear $\sigma$-model symmetry}
\MLreVV{Now that we have rationalized the amplitude of fluctuations $\sigma^2(E)$ and obtained an estimate of the energy-dependent dimensionless conductance $g(E)$, let us turn again to the detailed fractal structure of iso-energy LDOS maps.} We mentioned non-linear $\sigma$ models, a class of field theories inspired by the scaling theory of localization to describe low energy modes – diffusons and cooperons – of disordered conductors. Over a few decades, NL$\sigma$Ms have strongly advanced the understanding of disordered metals and superconductors as well as metal-insulator transitions. The formal mapping between supersymmetric sigma models and random matrix theory allowed relating scaling properties of the DOS moments with the symmetry class of the underlying tight-binding Hamiltonian, for instance whether spin-rotation, time reversal or chiral symmetry is preserved \cite{efetov1982supersymmetry,efetov1983supersymmetry,evers2008anderson}. Interestingly, the algebraic structure of $\sigma$-models themselves introduces another symmetry, the so-called Weyl-group symmetry, rooted in the algebraic structure of the field theory \cite{gruzberg2011symmetries}. This underlying symmetry, constrains the distribution functions and the multifractal spectra of the LDOS, in the form of exact relations \cite{gruzberg2011symmetries,gruzberg2013classification} which hold generally for very large classes of Hamiltonians -- including all Wigner-Dyson classes \cite{dyson1962statistical,gruzberg2013classification}. 
Let us quickly sketch the historical developments of these relations. In 1994, a symmetry relation for the LDOS distribution function at the metal-insulator transition was derived \cite{mirlin1994distribution,evers2008multifractality}.
\begin{equation}\label{symmetry_P_rho}
	\MLrev{P(\tilde{\rho})=\tilde{\rho}^{-3} P(\tilde{\rho}^{-1})}
\end{equation} followed a few years later by an analog relation for multifractal spectra \cite{mirlin2006exact}:
\begin{equation}\label{symmetry_MF}
f(\alpha) = f(2d-\alpha)-d+\alpha
\end{equation}
\MLrev{As discussed earlier, the multifractal spectrum $f(\alpha)$ describes the scaling exponents of the spatial LDOS distribution at a fixed energy. Eq.\ref{symmetry_MF} enforces a left-right symmetry on multifractal spectra, \textit{i.e.} a symmetry between regions of weak and strong deviations to the mean LDOS. It is a few years later that, as we mentioned, Eqs.(\ref{symmetry_P_rho}) and (\ref{symmetry_MF}) were shown to hold at the level of NL$\sigma$Ms by Gruzberg et al. \cite{gruzberg2011symmetries}, thus offering a powerful test of the validity of NL$\sigma$Ms, in particular for strongly disordered interacting systems.}

\MLrev{Indeed, since Finkel'stein introduced electron-electron interactions into NL$\sigma$Ms \cite{finkelstein1983frequency,burmistrov2019finkel}, the theory lead among other predictions, to multifractally-enhanced superconductivity in the strong interaction-strong disorder conditions despite the relevance of NL$\sigma$Ms being uncertain in this limit \cite{feigel2010fractal,burmistrov2012enhancement,andriyakhina2022multifractally}. In this context, the Weyl-group symmetry relations of Eqs.(\ref{symmetry_P_rho}) and (\ref{symmetry_MF}) emerge as a very powerful tool to test the validity of NL$\sigma$Ms and were recently tested numerically for the power-law banded random matrices models \cite{mildenberger2007boundary}, the 2D Anderson model in the symplectic class \cite{mildenberger2007wave,obuse2007multifractality} and the 3d Anderson transition \cite{vasquez2008multifractal}. To our knowledge, however, they were not yet tested on experimental datasets. In this context, we use Eq.\ref{symmetry_P_rho}-\ref{symmetry_MF} to test the presence of the fundamental Weyl-group symmetry of NL$\sigma$Ms in the n-doped Sn/Si correlated 2D system.}



\subsubsection*{Distribution functions}
Let us firstly turn to the distribution of normalized tunneling conductance : $P(\tilde{\eta})$ where $\tilde{\eta}=\eta/\langle \eta \rangle_r$. On Figure \ref{Figure4_symmetry}\textbf{a}, we plot the distribution functions $P(\tilde{\eta})$ of the normalized differential conductance dI/dV at various energies (solid line). We see that the LDOS distributions are continuously broadened when approaching the band edge. Notably, this broadening is very asymmetric and the distribution's fat-tail is strengthened when approaching the band-edge, \MLrev{in good agreement with recent measurements on Pb-Bi surface alloys \cite{jack2021visualizing}}.

To test Eq.\ref{symmetry_P_rho}, we plot on Figure \ref{Figure4_symmetry} the dual distribution \MLrev{$\tilde{\eta}^{-3} P(\tilde{\eta}^{-1})$} (dashed lines) which is predicted to match $P(\tilde{\eta})$. Clearly, the agreement is excellent in the metallic regime. \MLrev{For localized states ($E > E_{\rm c} = -0.65$ eV) however, a systematic difference between the solid and dashed lines develops until strong deviations to Eq.\ref{symmetry_P_rho} are obtained above $-0.5$ eV. We have to consider here that the exact symmetry relation Eq.\ref{symmetry_P_rho} is written for the density of states $\rho$ which differs by an unknown constant from the tunneling conductance and forces us to use the normalized DOS instead. At low disorder, Lerner has shown that normalized DOS has a log-normal distribution, which follows Eq.\ref{symmetry_P_rho} \cite{lerner1988distribution}. At higher disorder however, it is harder to exclude this effect and we cannot fully rule out that the deviations to Eq.\ref{symmetry_P_rho} are due to the normalization of the DOS.}





\subsubsection*{Multifractal spectra}

\begin{figure*}[!ht]
	\centering
	\includegraphics[width=\linewidth]{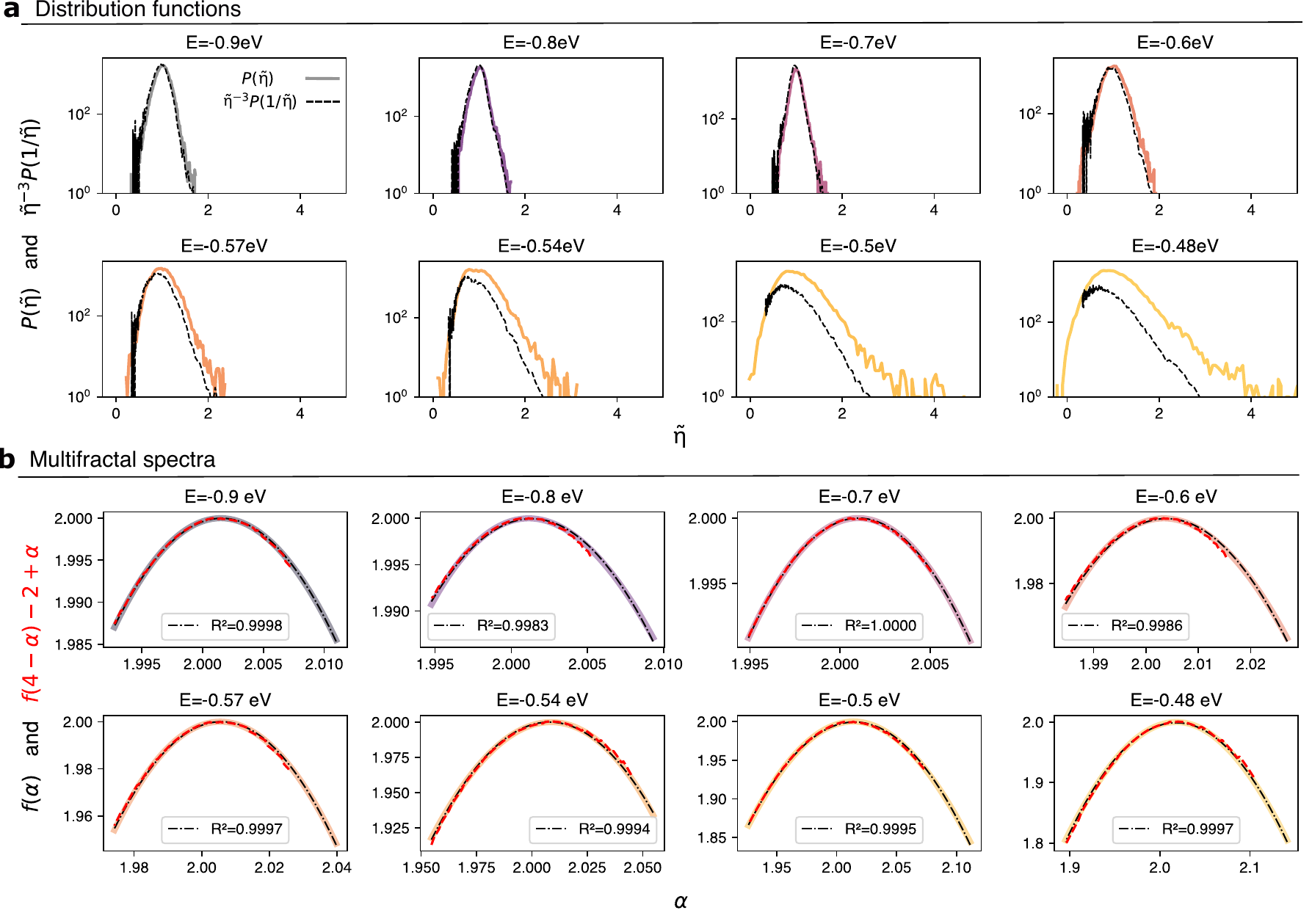} 
	\caption{\label{Figure4_symmetry}\MLeditor{\textbf{Distribution functions and multifractal analysis}} \textbf{a} Solid lines denote the dI/dV distribution functions \MLeditor{(or histogram) $P(\tilde{\eta})$ of differential conductance values} $\tilde{\eta}(E)$ at energy E = $\{-0.9, -0.8, -0.7, -0.6 ,-0.57, -0.54, -0.5, -0.48\}$ eV. Black dashed lines correspond to $\tilde{\eta}^{-3}P(1/\tilde{\eta})$. We recall that the mobility edge is at -0.65 eV. \textbf{b} Solid lines denote the multifractal spectra $f(\alpha)$ of dI/dV maps at energy E = $\{-0.9, -0.8, -0.7,-0.6,-0.57, -0.54, -0.5, -0.48\}$ eV. Red dashed lines account for $f(4-\alpha)-2+\alpha$. Multifractal spectra $f(\alpha)$ are fitted with the weak-multifractality solution Eq.\ref{Weak_MF_fit} (black dash-dotted lines). $\alpha_0$ is given in \MLeditor{Supplementary Note 1 and Supplementary Figure 2 along with a polynomial interpolation in Table S3.}}
\end{figure*}

On Figure \ref{Figure4_symmetry}\textbf{b}, we plot the multifractal spectra $f(\alpha)$ across the metal-insulator transition (already shown in Figure \ref{Correlation_heat_maps}\textbf{c}). Here $2(\alpha-2)$ can be seen as a scaling exponent and $f(\alpha)$ as the probability density of this scaling on a given map. When approaching the insulating phase, the spectra are broadened and shifted to the right, reaching their apex at $\alpha_0 > 2$ \cite{schreiber1991multifractal}. 

To test the symmetry relation Eq.\ref{symmetry_MF}, we plot with red dashed lines the right-hand side of Eq.\ref{symmetry_MF}, $f(4-\alpha)-2+\alpha$ on Figure \ref{Figure4_symmetry}\textbf{b}. Both spectra remain perfectly on top of each other down to -0.48 eV, only 0.1 V away from the true band-edge. This perfect agreement across the metal-insulator transition provides a very convincing proof of the validity of Eq.\ref{symmetry_MF} up to the strongly localized regime and in turn demonstrates the robustness of NL$\sigma$Ms.

\subsubsection*{Weak-multifractality fits}
\MLrev{Although multifractal spectra are not known theoretically at strong disorder, a single parameter controls the low-disorder spectrum : the multifractal exponent $\Delta_2$ (see $\Delta_2(E)$ on Figure \ref{Correlation_heat_maps}\textbf{c}), and $\Delta_2$ itself scales with inverse conductance $\Delta_2 \sim -1/g$, similarly to $\sigma^2$. In this limit, the spectra are peaked at $\alpha_0 = 2-\Delta_2/2$ and write
\begin{equation}\label{Weak_MF_fit}
f(\alpha) = 2-\frac{(\alpha-\alpha_0)^2}{4(\alpha_0-2)}	
\end{equation} We show on Figure \ref{Figure4_symmetry} that Eq.\ref{Weak_MF_fit} fits perfectly at all energies, with correlation factors $R^2$ above 0.995 (dash-dotted lines). We give a polynomial interpolation formula of $\alpha_0(E)$ in \MLeditor{Supplementary Note 1} to allow one to easily replot the full set of multifractal spectra.}

\MLrev{In surprising agreement with the weak-multifractality phenomenology, the full set of multifractal spectra across the metal-insulator transition is controlled by the energy-dependent multifractal exponent $\Delta_2(E)$ (plotted on Figure \ref{Correlation_heat_maps}\textbf{c}). This strong $f(\alpha)$ parabolicity down to $g\sim 0.1$ is a stronger result than the Weyl-group symmetry (which it directly implies) and fully supports the validity of NL$\sigma$Ms in the localized regime of the n-doped Sn/Si monolayer.}

\subsubsection*{Spectral correlation functions}
For the sake of completeness, we also computed the energy-energy correlation functions experimentally, and compared with theoretical predictions from the NL$\sigma$M (see \MLeditor{Supplementary Notes 2 \& 4 and Supplementary Figures 3 \& 7}).




\section*{Discussion}

\MLreV{Let us now draw the conclusions from our combined analysis of LDOS correlations, energy scalings of fluctuations and multifractal spectra. From spatial correlation functions, we found the mobility edge and the energy scaling of correlation length at the effective metal-insulator transition. Then, using $\sigma(E)$ scalings and a parabolic band Drude model, we determined that the dimensionless conductance $g(E)$ drops by more than one decade through the metal-insulator transition, from roughly 5 in the metallic range down to 0.1 at the band-edge. Finally, we demonstrated that the multifractal scalings of iso-energy LDOS maps closely follow a range of weak-disorder predictions from NL$\sigma$Ms, being eventually fully parametrized by the energy-dependent multifractal exponent $\Delta_2(E)$. In particular, the $f(\alpha)$ spectra obey Weyl-group symmetry relations based in NL$\sigma$Ms, thus supporting the validity of NL$\sigma$Ms in disordered and strongly correlated 2d electronic systems.}

\MLreV{A key point regarding Weyl-group symmetry relation Eq.(\ref{symmetry_MF}) is its holding on the Wigner-Dyson symmetry classes \cite{dyson1962statistical,gruzberg2011symmetries,gruzberg2013classification}, which includes the orthogonal class of non-interacting normal metals, the unitary class for non time-reversal symmetric systems and the symplectic classes when spin-rotational symmetry is lost. Interestingly, recent reports of chiral superconductivity in the hole-doped Sn/Si phase suggest that its ground-state may be in the chiral or Bogoliubov-de-Gennes classes instead of Wigner-Dyson \cite{ming2023evidence}. Thus, it would be very valuable to measure multifractal spectra in the highly p-doped samples down to the superconducting phase, possibly taking advantage of the vanishing DOS at the superconducting gap-edge to probe localized states \cite{lizee2023local}.}

\MLreV{Finally, we suggest extending our study to universal conductance fluctuations instead of LDOS maps, since they are measured from comparatively simpler magneto-transport measurements. Recently, enhanced multifractality at the Dirac peak of high-mobility graphene was reported \cite{amin2018exotic}. Systematic multifractal analysis of UCF from magneto-conductance measurements could then help to determine the symmetries of unknown quantum materials.}

\section*{Conclusion}
\MLrev{Building on recent experimental measurements of multifractality in tunneling conductance maps \cite{richardella2010visualizing,jack2021visualizing,rubio2020visualization,zhao2019disorder},} we map LDOS fluctuations through the metal-insulator transition at the valence band-edge of a strongly correlated 2d material, the $\sqrt{3}\times\sqrt{3}$ phase of tin on silicon. From LDOS spatial correlations, we find \MLL{an apparent mobility edge with a critical exponent $\nu_{\rm app} = 0.75$ for the correlation length} and report an energy scaling for the relative LDOS variance close to the band-edge $\sigma^2 \sim \vert E-E_{\rm edge} \vert^{-1.7}$ which is slightly modified by a 6.5 T transverse magnetic field. \MLrev{Using the relation between multifractal exponents and conductance derived from NL$\sigma$Ms, we estimate the energy-dependent dimensionless conductance $g$ across the metal-insulator transition and explain the strong increase in LDOS relative variance at the band-edge.}
\MLrev{Then, we successfully test multifractal symmetry relations based on the algebraic structure of NL$\sigma$Ms of the Wigner-Dyson symmetry class. We show that even on such a strongly correlated system, they hold up from $g\sim 5$ down to at least $g \sim 0.1$, thus constraining theories for highly-disordered interacting systems. We further show that multifractal spectra in fact obey a stronger law down to $g\sim0.1$, the weak-disorder parabolic form derived from NL$\sigma$Ms.}

To help unifying localization studies on correlated electronic systems, we bridge together many key results in a consistent picture of a single model material. \MLrev{Our findings pave the way to infer electronic symmetries from quantum interference signals. If the p-doped Sn/Si monolayer is a chiral superconductor, as recently reported \cite{ming2023evidence} -- then multifractal spectra should not be of Wigner-Dyson symmetry and could obey different symmetry relations \cite{gruzberg2013classification}. Finally, multifractal scalings in the universal conductance fluctuations should also show Weyl-group symmetry relations \cite{gruzberg2013classification} and recent measurements in high-mobility graphene strongly suggest extending our methodology to transport experiments \cite{amin2018exotic}.}

\section*{Acknowledgements}
We dedicate this work to the memory of our dear colleague and friend Mark-Oliver Goerbig, who left us far too soon. Mark was a passionate mesoscopic physicist and a brilliant professor, whose kindness, enthusiasm, and profound physical insight will be dearly missed.

\MLeditor{
\section*{Methods}
\subsection*{Sample preparation}
In the preparation chamber of our homemade scanning-tunneling microscope, we prepared the $\sqrt{3}\times\sqrt{3}$ reconstruction of tin on silicon (111) following the standard recipe \cite{modesti2007insulating,odobescu2017electronic,li2013magnetic}: a highly n-doped Si(111) sample with room-temperature resistivity in the m$\Omega$.cm range was flashed to 1100$^{\circ}$C in ultra-high vacuum and slowly cooled down to 600$^{\circ}$C until the 7$\times$7 reconstruction was consistently observed. Keeping the substrate's temperature at 600$^{\circ}$C, we then evaporated 1/3 monolayer of tin from an e-beam evaporator (at a rate of about 0.15 monolayer per min) before cooling down the sample to 500$^{\circ}$C in one minute.}
\MLeditor{
\subsection*{STM measurements}
We measured the sample at 300 mK in a homemade STM system with base pressure of $10^{-11}$ mbar. We use metallic PtIr tips. At each point of a 50 nm square grid, we set the tip height using a current setpoint of 100 pA under a bias voltage of -1V and record the I(V) spectrum up to -0.4V. The differential conductance is then computed numerically. The magnetic field is applied perpendicular to the sample's surface with superconducting coils. Additional STM maps can be found in Supplementary Methods.}

\MLeditor{
\section*{Data availability}
\MLeditoR{The figures can be replotted from the Supplementary data file. The original data} that support the findings of this study are available from the corresponding authors (Mathieu Lizée or Tristan Cren) upon reasonable request.}

\MLeditor{
\section*{Author contributions}
M.L, M.T, F.D, M.H, C.B and T.C performed the experiments. M.L, C.B and T.C conceived the project. M.L analyzed the data and wrote the manuscript. I.B provided theoretical support. All authors discussed the results and contributed to the manuscript.}

\MLeditoR{\section*{Competing interests}
 The authors declare no competing interests.}

\section*{References}
\bibliography{Sn_Si.bib}
\end{document}